\documentclass[pre,twocolumn,showpacs]{revtex4-1}
\usepackage{graphicx}
\usepackage{amsmath,amssymb}

\newcommand{\latin}[1]{{\itshape #1}}

\newcommand{\ie}{\latin{i.\,e.}}

\begin{document}

\title{Revisiting Riccioli's free fall calculations}

\author{Patrick B. Warren}

\email{patrick.warren@unilever.com}

\affiliation{Unilever R\&D Port Sunlight, Quarry Road East, Bebington,
  Wirral, CH63 3JW, UK.}

\date{March 19, 2013}

\begin{abstract}
In 1651 Giovanni Riccioli reported the earliest accurate measurements
the acceleration due to gravity, $g$, from pendulum-timed free fall
experiments.  The use of Huygens' pendulum formula (published 1673)
allows one to deduce the pendulum length from this data, free from
assumptions about the conversion to modern units, and independent of
the actual value of $g$.  When this length is compared to the reported
pendulum length, a 15\% systematic error is revealed.  This could
perhaps be attributed to the difficulty Riccioli faced in subdividing
the contemporary unit of length (the Roman foot) to the requisite
millimeter accuracy.
\end{abstract}

\pacs{%
01.65.+g, 
45.40.Gj, 
91.10.Pp} 

\maketitle

In 1651 Giovanni Riccioli published measurements of the acceleration
due to gravity from pendulum-timed free fall experiments, in his
\emph{Almagestum Novum} (a treatise on mathematics and astronomy,
updating Ptolemy's 2nd century \emph{Almagest}).  Riccioli confirmed
Galileo's observation reported twenty years previously (in 1632) that
the distance $z$ fallen in a time $t$ satisfies $z\propto t^2$.
Riccioli's measurements were the most accurate to that date of the
constant of proportionality, which in modern terminology is $g/2$.  A
clear-cut comparison with the current accepted value $g\approx9.81
\mathrm{m}\,\mathrm{s}^{-2}$ is frustrated by uncertainty over the
conversion of Riccioli's unit of length, the Roman foot, to modern
units.  All this has been elegantly explained in an article in Physics
Today by Graney \cite{Gra12}, who also provides a modern translation
of the relevant part of the \emph{Almagestum Novum} \cite{Gra12b}.
Graney's article was the inspiration for this short note, which
extends the present author's recently published Letter in Physics
Today \cite{War13}.

\begin{figure}
\begin{center}
\includegraphics[clip=true,width=2.5in]{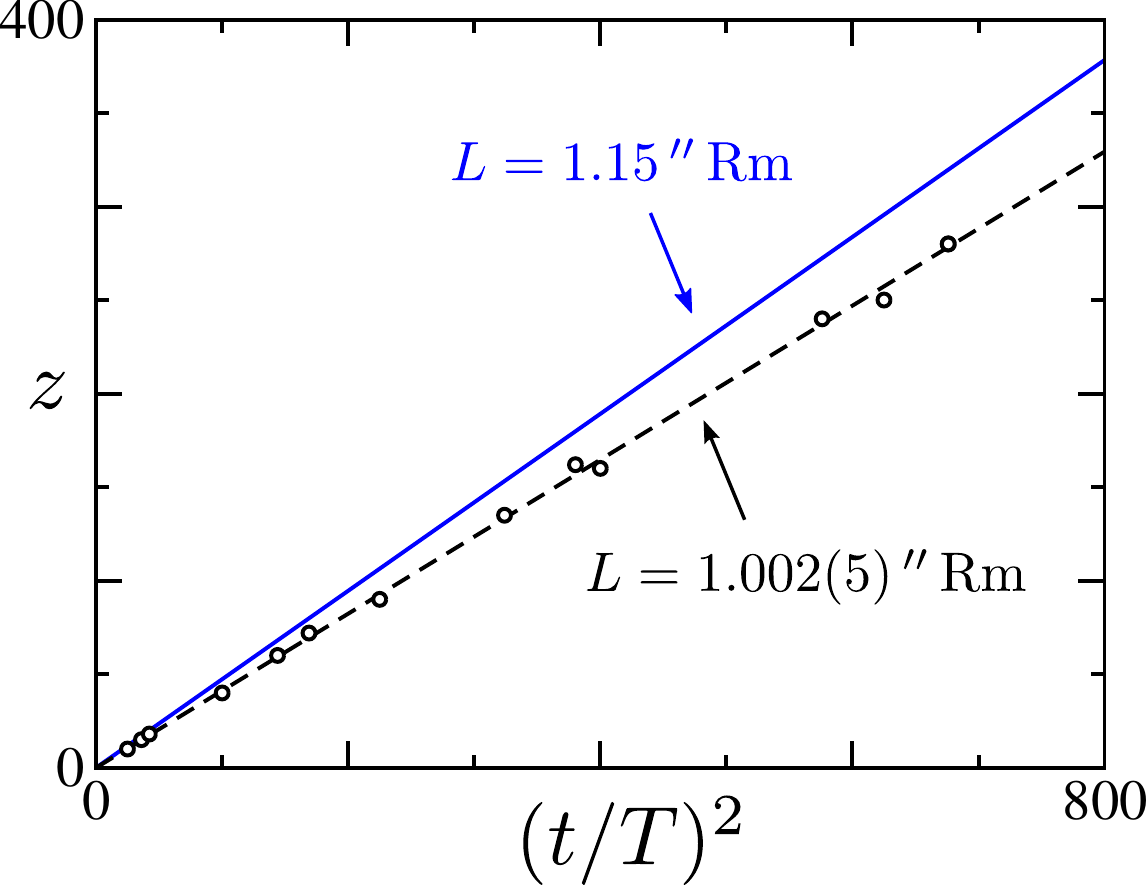}
\end{center}
\vskip -0.5cm
\caption{A plot of the free fall distance as a function of the square
  of the free fall time measured in pendulum half periods should obey
  Eq.~\eqref{eq:1}.  The circles are Riccioli's data, with a best-fit
  line (dashed).  The solid line (blue) is the prediction using
  Riccioli's reported pendulum length.
\label{fig:plot}}
\end{figure}

Riccioli's actual free fall data is shown in Table \ref{tab:data}.  The
second column in the Table is the free fall time, measured in pendulum
half periods (approximately 1/6 second).  The fifth column in the
Table is the free fall distance, measured in Roman Feet.
Unbeknownst to Riccioli at the time, but known to us thanks to
Christiaan Huygens, the half period of a pendulum is $T = \pi \sqrt{L
  / g}$ where $L$ is the length of the pendulum.  This can be combined
with the free fall law, $z=gt^2/2$, to get 
\begin{equation}
z = \frac{\pi^2L}{2} \times \Bigl(\frac{t}{T}\Bigr)^2\,.\label{eq:1}
\end{equation}
So a plot of the free fall distance against the square of the free
fall time measured in pendulum half periods should be a straight line
through the origin with a slope $\pi^2L / 2$.  Hence we can work out
the length of the pendulum, at least in Roman Feet.  Riccioli's data
plotted in this way is shown in Fig.~\ref{fig:plot}: it is a very good
fit to a straight line.  The best fit line (through the origin) has
slope $\pi^2L / 2 = 0.412(2)$ where the figure in brackets is the
reported least squares fitting error in the final digit.  Hence $L =
0.0834(4)$ Roman feet, or $1.002(5)$ Roman inches.

\begin{table}
\begin{center}
\includegraphics[clip=true,width=3.4in]{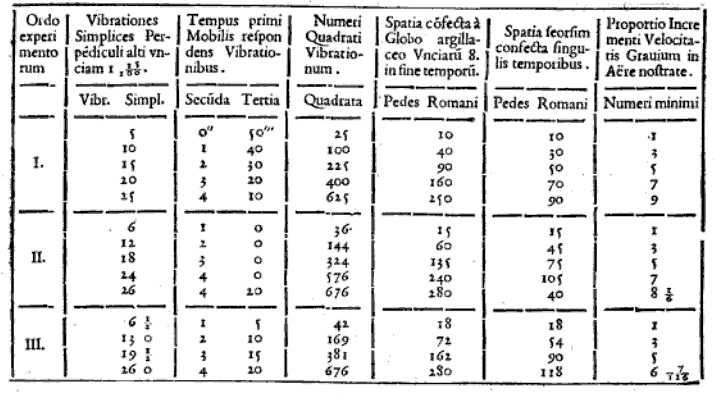}
\end{center}
\vskip -0.5cm
\caption{Riccioli's original data from the \emph{Almagestum Novum}
  (p387).  The table was taken from Google Books.  \label{tab:data}}
\end{table}

This can be compared with Riccioli's report of the pendulum length
which ``measured to the center of the little bob is one and fifteen
hundredths of the twelfth part of an old Roman foot'' \cite{Gra12b}.
So Riccioli has $L= 1.15$ Roman inches (with an inferred accuracy
of 5 parts in 100).  The prediction from this is shown as the solid
(blue) line in Fig.~\ref{fig:plot}.  Clearly, a systematic error has
crept in somewhere.  Moreover, the discrepancy is independent of the
conversion to modern units, and independent of the actual value of $g$,
which cancels in Eq.~\eqref{eq:1}.

The data is consistent with the pendulum being 15\% shorter than
Riccioli claimed, or equivalently the oscillation frequency being 7\%
too rapid (recall $T\sim L^{1/2}$).  However the discrepancy is in the
wrong direction to be explained either by the fact (known to Riccioli
by that time) that large amplitude pendulum swings are no longer
isochronous (otherwise the pendulum would be slower than expected) or
that it should properly be regarded as a compound pendulum whose
centre of oscillation lies below the centre of mass (again it would be
slower than expected).  Unless there is a typographical error, the
most plausible explanation seems to lie with Riccioli's measurement of
the pendulum length, perhaps in the difficulty in subdividing a Roman
foot into 5/100 parts of Roman inches (\ie\ 240 subdivisions).  This
is an accuracy of about $1\,\mathrm{mm}$ in modern units so a
systematic error seems perfectly excusable.

Let me end with a philosophical digression.  What entitles us to
conclude there is a systematic error?  The answer is that it doesn't
agree with Huygens' theory for the period of a pendulum.  But surely
this is exactly backwards?  Experiment is supposed to confirm theory,
not theory disprove experiment!  Actually in practised science this is
much more common than might be expected, and has even been elevated to
a principle.  The precise quote, attributed to Arthur Eddington, is :
\begin{quote} 
\itshape It is also a good rule not to put over much confidence in the
observational results that are put forward until they are confirmed by
theory.
\end{quote}
Galileo had something to say on systematic errors \cite{CS38},
discussing deviations from his observation that all objects
fall at the same rate :
\begin{quote} 
\itshape But, Simplicio, I trust you will not follow the example of
many others who divert the discussion from its main intent and fasten
upon some statement of mine which lacks a hair’s-breadth of the truth
and, under this hair, hide the fault of another which is as big as a
ship's cable.
\end{quote}
Physics is an experimental science and the nitty-gritty is largely
about the control and elimination of sources of systematic error.
Knowing when to stop, and awareness that ``perfect is the enemy of
  good'' \cite{Vol72}, is I guess the mark of a good experimentalist.

For context, here's a brief chronology of related works published in
the 17th century :
\begin{itemize}
\item 1632 -- Galileo -- \emph{Dialogue},
\item 1651 -- Riccioli -- \emph{Almagestum Novum},
\item 1673 -- Huygens -- \emph{Horologium Oscillatorium},
\item 1687 -- Newton -- \emph{Principia}.
\end{itemize}

\end{document}